\documentclass[final,5p,times,twocolumn]{elsarticle}

\newcommand{\ice}[1]{\relax}

\usepackage{graphicx,amsmath,amssymb}

\newcommand{\myrm}{\relax}

\begin{document}

\begin{frontmatter}

\title{
\vspace*{-1cm}
\boldmath \bf
Adler Function, Sum Rules and Crewther Relation of Order  ${\cal O}(\alpha_s^4)$:
 the Singlet Case}

\author[a]{P.~A.~Baikov}
\author[b]{K. G. Chetyrkin}
\author[b]{J.~H.~K\"uhn}
\author[b]{J.~Rittinger}

\address[a]{
Skobeltsyn Institute of Nuclear Physics, Lomonosov Moscow State University, 
1(2), Leninskie gory, Moscow 119234, Russian Federation
        }
\address[b]{Institut f\"ur Theoretische Teilchenphysik, Karlsruhe
  Institute of Technology (KIT), Wolfgang-Gaede-Stra\ss{}e 1, 726128 Karlsruhe, Germany}
\begin{abstract}
The analytic result for the singlet part of the Adler function of the
vector current in a general gauge theory is presented in five-loop
approximation.  Comparing this result with the corresponding singlet
part of the Gross-Llewellyn Smith sum rule \cite{Baikov:2010iw}, we successfully
demonstrate the validity of the generalized Crewther relation for the
singlet part. This provides  a  non-trivial test of both
our calculations and the generalized Crewther relation.
Combining the result    with the already available non-singlet part of the Adler function 
\cite{Baikov:2008jh,PhysRevLett.104.132004}
we  arrive at the complete  ${\cal O}(\alpha_s^4)$  expression for the Adler 
function and, as  a direct consequence, at  the complete  ${\cal O}(\alpha_s^4)$  
correction to the $e^+ e^-$ annihilation into hadrons in a general gauge theory. 
\end{abstract}

\end{frontmatter}

\section{Introduction}
\label{sec:intro}

The  classical Crewther relation (CR) \cite{Crewther:1972kn}
connects in a non-trivial way two seemingly unrelated quantities,
namely the Adler function \cite{Adler:1974gd} and perturbative
corrections arising in the sum rules relevant for deep inelastic
scattering (DIS). Originally the CR had been formulated for the case
of a conformal-invariant limit of a field   theory.  Subsequently, observing a
close relation between the ${\cal O}(\alpha_s^3)$ terms in the Adler
function and the corrections to the Bjorken sum rule for polarized electron-nucleon scattering, its
generalization for the case of QCD was suggested in
\cite{Broadhurst:1993ru}, introducing as modification additional terms
proportional to the $beta$-function. More formal arguments for
the validity of this ``generalized Crewther relation'' (GCR) were given
in \cite{Gabadadze:1995ei,Crewther:1997ux,Braun:2003rp}.  During the past years the
perturbative corrections both for Adler function and Bjorken sum rule
were extended from ${\cal O}(\alpha_s^3)$
\cite{Gorishnii:1990vf,Surguladze:1990tg,Chetyrkin:1996ez,Larin:1991tj}
to ${\cal O}(\alpha_s^4)$
\cite{Baikov:2008jh,PhysRevLett.104.132004}. 
However, these results were restricted to the respective non-singlet
parts. Nevertheless, they could be used to demonstrate the validity of
the GCR between non-singlet Adler function and Bjorken sum rule
\cite{Bjorken:1967px,Bjorken:1969mm}, thus providing at the same time
an important cross check of the underlying, demanding calculations.

The ${\cal O}(\alpha_s^4)$ singlet piece of the sum rule was published
in \cite{Baikov:2010iw}, thus completing the prediction for the
Gross-Llewellyn Smith (GLS) sum rule \cite{Gross:1969jf}. Below we
give the corresponding result for the Adler function. On the one hand
this leads to a prediction of the familiar $R$-ratio measured in
electron-positron annihilation, including the (small) up to now missing
singlet pieces of ${\cal O}(\alpha_s^4)$, on the other hand this
result allows to test the GCR also for the singlet case.
Note that all results discussed in this paper are  assumed to be  renormalized within the 
conventional $\overline{\mbox{MS}}$ subtraction scheme  \cite{Bardeen:1978yd}.

\section{Singlet ${\cal O}(\alpha_s^4)$ contributions to the Adler function and R(s)}

For the definition of the Adler function it is convenient to start with the polarization  function of the flavor singlet vector current:
\begin{equation}
3 \, Q^2 \, \Pi(Q^2) =
i\,\int{\rm d}^4 x \, e^{iq\cdot x}\langle 0|{\rm  T}j_\mu(x)j^\mu(0)|0\rangle~,
\end{equation}
with $j_{\mu} = \sum_i\overline{\psi}_i\gamma_{\mu}\psi_i$ and  $Q^2 = -q^2$. The  corresponding Adler function
\begin{equation}
D(Q^2) =  -12\, \pi^2 Q^2\, \frac{\rm{d}}{\rm{d} Q^2} \Pi(Q^2)
\end{equation}
is naturally decomposed into a sum of the non-singlet (NS) and singlet (SI) components (see Fig.~1):
\begin{equation}
D(Q^2) = n_f\, \,  D^{\myrm NS}(Q^2) + n_f^2 \, D^{\myrm SI}(Q^2)~.\label{D:decomp}
\end{equation}
Here $n_f$ stands for   the total number of quark flavours; all quarks  are  considered as massless.
%Similar decomposition holds for the very polarization operator.
\begin{figure*}
\begin{center}
\includegraphics[width=10cm]{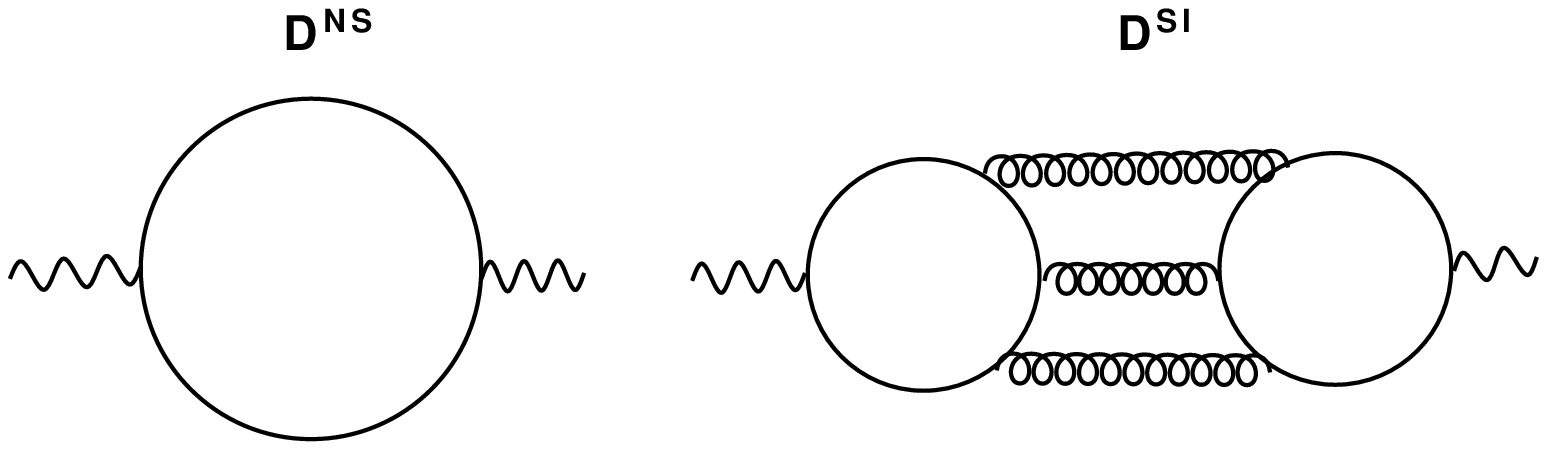}
\end{center}
\hspace{6cm} {\bf (a)}  \hspace{6cm} {\bf (b)}
\caption{
Lowest order non-singlet (a) and singlet (b) diagrams contributing to the Adler function.
}
\label{figure:1}
\end{figure*}
The Adler function $D^{\myrm EM}$ corresponding to the electromagnetic vector current $j^{\myrm EM}_{\mu} = \sum_i q_i\, \overline{\psi}_i\gamma_{\mu}\psi_i$ ($q_i$ stands for the electric charge of the quark field $\psi_i$)  is thus  given by the following combination:
\begin{equation}
D^{\myrm EM} = \bigg(  \sum_i q_i^2 \bigg) D^{\myrm NS} + \bigg(  \sum_i q_i \bigg)^2 \, D^{\myrm SI}~.
\end{equation}
Similar decompositions hold for the  corresponding polarization functions  $\Pi$ and $\Pi^{\myrm EM}$.

The  physical observable $ R(s) ={\sigma(e^+e^-\to {\rm hadrons})\over \sigma(e^+e^-\to \mu^+\mu^-)}$ is related to $\Pi^{\myrm EM}(Q^2)$ by the optical theorem
\begin{equation}
R(s) = 12 \pi \,\Im \, \Pi^{\myrm EM}(-s- i\epsilon)~.
\label{R(S):def}
\end{equation}
The result for the perturbative expansions of  the non-singlet part ($a_s \equiv \frac{\alpha_s}{\pi}$)
\begin{equation}
{D}^{\myrm NS}(Q^2) = d_R\,\bigg( 1 + \sum_{i=1}^{\infty} \ d^{\myrm NS}_i \, a_s^i(Q^2) \bigg)
\end{equation}
has been presented in \cite{PhysRevLett.104.132004}. For the  singlet 
%Adler functions 
part it reads:
\begin{equation}
D^{\myrm SI}(Q^2) =  d_R\,\bigg( \sum_{i=3}^{\infty} \  d^{\myrm SI}_i \, a_s^i(Q^2) \bigg)~,
\end{equation}
where the parameter $d_R$ (the dimension of the quark color representation, $d_R= 3$ in QCD) is  factorized in {\em both} non-singlet and singlet components. 

The singlet component has the following structure at orders $\alpha_s^3$ and $\alpha_s^4$:
\begin{eqnarray}
d_3^{\myrm SI} &=& d^{abc}\,d^{abc}/d_R\,\Big( \tfrac{11}{192} - \tfrac{1}{8}\,\zeta_{3}\Big)~,
\\ 
d_4^{\myrm SI} &=& d^{abc}\,d^{abc}/d_R\,\Big( C_F \, d_{4,1}^{\myrm SI} +C_A \, d_{4,2}^{\myrm SI} + T\, n_f\, d_{4,3}^{\myrm SI}\Big)~.
\label{SI:d3:d4}
\end{eqnarray}
Here  $C_F$ and $ C_A$ are the quadratic Casimir operators of the fundamental and the adjoint representation of the Lie algebra, $d^{abc} = 2\, \mathrm{Tr}(\{\frac{\lambda^a}{2},\frac{\lambda^b}{2}\},  \frac{\lambda^c}{2}\})$, $T$ is the trace normalization of the fundamental representation. 
% \ice{
% The exact definitions of
% ${d_F^{a b c } d_F^{a b c }}$
% are given in \cite{Vermaseren:1997fq}.
% }
For  QCD (colour gauge group SU(3)):
\begin{equation*}
C_F =4/3~,~~~ C_A=3~,~~~ T=1/2~,~~~ {d^{a b c }\,d^{a b c }} = 40/3~.
\end{equation*}

% \ice{
% The evaluation of the NLO terms of $r_V^{\myrm S}$ requires the
% calculation of the absorptive parts of five-loop diagrams with
% massless propagators which, with the help of some complicated
% combinatorics based on the $R^*$-operation \cite{Chetyrkin:1984xa}, can be boiled down
% to the calculation of  four-loop propagator
% diagrams. The  latter have been computed  
% via reduction to 28  master integrals, based on evaluating  
% sufficiently many terms of the $1/D$ expansion \cite{Baikov:2005nv} of
% the corresponding coefficient functions \cite{Baikov:1996rk}. 
% This direct procedure required huge computing resources and
% was performed using a parallel version \cite{Tentyukov:2004hz} of FORM
% \cite{Vermaseren:2000nd}. The master integrals are reliably  known from
% \cite{Baikov:2010hf,Smirnov:2010hd,Lee:2011jt}. 
% The details of the calculation, the results in analytic form and their 
% relation to the Gross-Llewellyn Smith sum rule will be given in  \cite{Baikov:2012}.
% }
Using the methods described  in  \cite{Chetyrkin:1984xa,Baikov:2005nv,Baikov:1996rk,Baikov:2008jh,Baikov:2010hf,PhysRevLett.104.132004} we obtain
\begin{eqnarray}
d_{4,1}^{\myrm SI} &=& -\tfrac{13}{64} - \tfrac{1}{4}\,\zeta_3  + \tfrac{5}{8}\,\zeta_5~, 
\label{d41}
\\
d_{4,2}^{\myrm SI} &=& \tfrac{ 3893}{4608} - \tfrac{169}{128}\,\zeta_3 + \tfrac{45}{64}\,\zeta_5  -\tfrac{11}{32}\,\zeta_3^2~,
\label{d42}
\\
d_{4,3}^{\myrm SI} &=& -\tfrac{149}{576} + \tfrac{13}{32}\,\zeta_3 - \tfrac{5}{16}\,\zeta_5 + \tfrac{1}{8}\,\zeta_3^2~.
\label{d43}
\end{eqnarray}

With the use of eqs.~(\ref{d41})-(\ref{d43}) and the result for $D^{\myrm NS}$ from \cite{PhysRevLett.104.132004} we arrive at  the complete  result for the ratio $R(s)$ at  order $\alpha_s^4$ in (massless)  QCD:
\begin{align}
 R(s) =&~ 3\sum_f q_f^2 \bigg\{
1 + a_s + a_s^2 \Big( \tfrac{365}{24} - 11\,\zeta_3 - \tfrac{11}{12}\,n_f + \tfrac{2}{3}\,\zeta_3 \,n_f \Big)
\nonumber\\[-1mm]
&~+ a_s^3\,\Big[n_f^2\,\Big( \tfrac{151}{162} - \tfrac{1}{108}  \pi^2 - \tfrac{19}{27}\,\zeta_{3}\Big)
\nonumber \\
&\qquad  + n_f\,\Big( - \tfrac{7847}{216} + \tfrac{11}{36}\,\pi^2 + \tfrac{262}{9}\,\zeta_{3} - \tfrac{25}{9}\,\zeta_{5}\Big)
\nonumber \\
&\qquad + \tfrac{87029}{288} - \tfrac{121}{48}\,\pi^2 - \tfrac{1103}{4}\,\zeta_{3} + \tfrac{275}{6}\,\zeta_{5} \Big]
\nonumber \\
&~+ a_s^4\,\Big[ n_f^3\,\Big( - \tfrac{6131}{5832} + \tfrac{11}{432}\,\pi^2 + \tfrac{203}{324}\,\zeta_{3} - \tfrac{1}{54}\,\pi^2\,\zeta_{3}
  + \tfrac{5}{18}\,\zeta_{5}\Big)
\nonumber\\
&\qquad + n_f^2\,\Big( \tfrac{1045381}{15552} - \tfrac{593}{432}\,\pi^2 - \tfrac{40655}{864}\,\zeta_{3}
\nonumber\\
&\qquad\qquad\qquad + \tfrac{11}{12}\,\pi^2\,\zeta_{3}  + \tfrac{5}{6}\,\zeta_3^2 - \tfrac{260}{27}\,\zeta_{5}\Big)
\nonumber\\
&\qquad + n_f\,\Big( - \tfrac{13044007}{10368} + \tfrac{2263}{96}\,\pi^2 + \tfrac{12205}{12}\,\zeta_{3} - \tfrac{121}{8}\,\pi^2\,\zeta_{3}
\nonumber\\
&\qquad\qquad\qquad - 55\,\zeta_3^2 + \tfrac{29675}{432}\,\zeta_{5} + \tfrac{665}{72}\,\zeta_{7}\Big)
\nonumber\\
&\qquad + \tfrac{144939499}{20736} - \tfrac{49775}{384}\,\pi^2 - \tfrac{5693495}{864}\,\zeta_{3} + \tfrac{1331}{16}\,\pi^2\,\zeta_{3}
\nonumber\\
&\qquad\qquad\qquad  +\tfrac{5445}{8}\,\zeta_3^2 + \tfrac{65945}{288}\,\zeta_{5} - \tfrac{7315}{48}\,\zeta_{7}\Big]
\bigg\}
\nonumber
\end{align}
\begin{align}
&~ + \bigg(\sum_f q_f \bigg)^2 \bigg\{ a_s^3 \Big( \tfrac{55}{72} - \tfrac{5}{3}\,\zeta_3 \Big)
\nonumber\\[-2mm]
&\qquad\qquad + a_s^4\,\Big[ n_f\,\Big( - \tfrac{745}{432} + \tfrac{65}{24}\,\zeta_{3} + \tfrac{5}{6}\,\zeta_3^2 - \tfrac{25}{12}\,\zeta_{5}\Big)
\nonumber\\
&\qquad\qquad\qquad + \Big( \tfrac{5795}{192} - \tfrac{8245}{144}\,\zeta_{3} - \tfrac{55}{4}\,\zeta_3^2
  + \tfrac{2825}{72}\,\zeta_{5}\Big)\Big]\bigg\}~,
\label{Rqcd}
\end{align}
where we set $\mu=Q$. The full results for Adler function and $R(s)$  for generic color factors
and  generic value of $\mu$ are  rather lenghty and can be found  
available (in computer-readable form) in
{\tt http://www-ttp.physik.uni-karlsruhe.de/Progdata\\/ttp12/ttp12-017.} Numerically, it reads:
\begin{align}
R(s) =&~ 3\sum_f q_f^2\,\bigg\{
1 + a_s + a_s^2\,\Big( 1.986 - 0.1153 \,n_f \Big)
\nonumber\\[-2mm]
&\qquad+ a_s^3\,\Big( -6.637 - 1.200 \,n_f - 0.00518 \,n_f^2 \Big)
\nonumber\\
&\qquad +a_s^4\,\Big( -156.608 + 18.7748\,n_f  - 0.797434 \,n_f^2
\nonumber\\
&\qquad\qquad + 0.0215161 \,n_f^3 \Big)\bigg\}
\nonumber
\\
&~- \bigg(\sum_f q_f\bigg)^2 \Big( 1.2395\, a_s^3   + \Big(17.8277 - 0.57489 \,n_f \Big)\, a_s^4 \Big)~.
\nonumber
\end{align}
% \ice{
% \mbox{for $\,n_f$=5}  
% \[
% \frac{11}{3}
% \left[
% 1 + a_s  + a_s^2\, 1.409 
% -12.767\, a_s^3  - 79.98\, a_s^4 
% \right]
% + \frac{1}{9}
% \left[
% -1.240\, a_s^3 {- 14.95\, a_s^4 }
% \right]
% \]
% }

Specifically, for the  particular values of $n_f = 4 $ and 5 one  obtains (for the terms of order $\alpha_s^3$ and  $\alpha_s^4$ we have explicitly decomposed the coefficient into  non-singlet and singlet contributions):
\begin{align}
R^{n_f=4}(s)  =&~ \frac{10}{3} \bigg[ 1 + a_s + 1.5245 a_s^2
\nonumber \\
&~+ a_s^3\, \Big( -11.686 =-11.52 - 0.16527^{\myrm SI}\Big)
\nonumber \\
&~+ a_s^4\, \Big(  - 94.961 = -92.891 - 2.0703^{\myrm SI}\Big) \bigg]~,
\\
R^{n_f=5}(s)  =&~ \frac{11}{3} \bigg[ 1 + a_s + 1.40902 a_s^2
\nonumber \\
&~+ a_s^3\, \Big( - 12.80=-12.767 - 0.037562^{\myrm SI}\Big)
\nonumber \\
&~+ a_s^4\, \Big(  - 80.434 = -79.981 - 0.4531^{\myrm SI}\Big) 
\bigg]
{}.
\end{align}
Note that for $n_f=3$ the singlet contributions vanish in every order in $\alpha_s$ as the corresponding
global coefficient $(\sum_i q_i)^2$ happens to be zero. 
Implications of this result for the determination of $\alpha_s$ in 
electron-positron annihilation and in $Z$-boson decays are discussed in
\cite{Baikov:2012er}.

\section{GLS sum rule at order ${\cal O}(\alpha_s^4)$ and the Crewther relation}
The second quantity of interest, the GLS sum rule,
\begin{equation}
\frac{1}{2}\int_0^1 F_3(x,Q^2) dx = 3\,  C^{CLS}(a_s) 
{},
\end{equation}
 relates the lowest  moment of 
the  isospin singlet structure function $F_3^{\nu p + \bar\nu p}(x,Q^2)$ to a coefficient 
$C^{CLS}(a_s)$, 
which appears in the operator product expansion of the axial and vector
non-singlet  currents
\begin{equation}
i\int T{A_{\mu}^{a}(x)V_{\nu}^{b}(0)}e^{iqx}dx
|_{\normalsize q^2\rightarrow -\infty}
\approx C_{\mu\nu\alpha}^{V,ab}\,V_{\alpha}(0)+ \dots
\label{OPE}
\end{equation}
where 
\[C_{\mu\nu\alpha}^{V,ab}
=
\delta^{ab}\epsilon_{\mu\nu\alpha\beta}
\frac{q^{\beta}}{Q^2}C^{GLS}(a_s)
{}
\]
and
$V_{\alpha} =\overline{\psi}\gamma_{\alpha} \psi$ is a flavour singlet  quark current.
At last 
$A_{\mu}^{a} = \overline{\psi}\gamma_{\mu} \gamma_5 t^a \psi$,  
$V_{\nu}^{b} = \overline{\psi}\gamma_{\nu} t^b \psi$ are axial vector    and  vector 
non-singlet quark currents, with  $t^a, \, t^b$  being the generators of the flavour group
$SU(n_f)$.

\begin{figure*}
\begin{center}
\includegraphics[width=12cm]{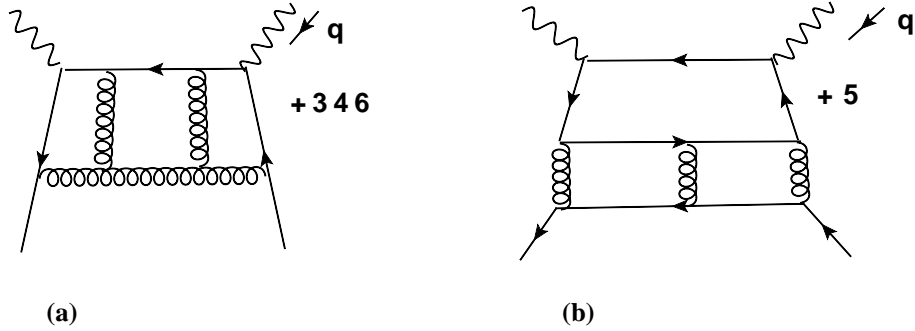}
\end{center}
\hspace{3.6cm} {\bf (a)}  \hspace{6.2cm} {\bf (b)}
\caption{
(a),(b): ${\cal O}(\alpha_s^3)$ non-singlet and singlet diagrams contributing to the Gross-Llewellyn Smith sum rule;
note that the  coefficient function  $C^{Bjp}$ is contributed by only non-singlet diagrams.
}
\label{table:1}
\end{figure*}

Again diagrams contributing to $C^{GLS}(a_s)$ can be separated in two groups:
non-singlet and singlet ones (see Fig. 2):
\begin{align}
C^{GLS} =&~ C^{\myrm NS} + C^{\myrm SI}~,
\label{GLSdecomp}
\\
C^{\myrm NS}(Q^2) =&~ 
1+
\sum_{i=1}^{\infty} \  c^{\myrm NS}_i \, a_s^i(Q^2)~,
\\
{C}^{\myrm SI}(Q^2) =&~
\sum_{i=3}^{\infty} \  c^{\myrm SI}_i \, a_s^i(Q^2)~.
\end{align}

The results for both functions  $C^{NS}$ and $C^{SI}$ at order 
$\alpha_s^3$ are known since early 90-ties \cite{Larin:1991tj}. 
Note that as a consequence of  chiral invariance the closely related
Bjorken sum rule receives contributions  from   the non-singlet piece only  \cite{Larin:1991tj}: 
\begin{equation}
C^{Bjp} \equiv C^{\myrm NS}
\label{chiral:inv}
{}.
\end{equation}

The ${\cal O}(\alpha_s^4)$ contribution to $C^{Bjp}$  has been  computed  some  time ago
\cite{PhysRevLett.104.132004}. 
The calculation of the  ${\cal O}(\alpha_s^4)$ contribution to  ${C}^{\myrm SI}$
has been  published in  \cite{Baikov:2010iw} for
 a  generic gauge group and is repeated below:
\begin{align}
c_3^{\myrm SI} =&~ n_f\, \frac{d^{abc}\,d^{abc}}{d_R} \left(
c_{3,1}^{\myrm SI} \equiv
-\frac{11}{192}
+\frac{1}{8}  \, \zeta_{3}
\right)~,
\label{c31SI}
\\
c_4^{\myrm SI} =&~
n_f\,\frac{d^{abc}\,d^{abc}}{d_R} \left(
C_F \,c_{4,1}^{\myrm SI} +C_A \, c_{4,2}^{\myrm SI} + T\, n_f\, c_{4,3}^{\myrm SI}
\right)~,
\label{c4SI}
\\
 c_{4,1}^{\myrm SI} =&~ {
\frac{37}{128}
+\frac{1}{16}  \, \zeta_{3}
-\frac{5}{8}  \, \zeta_{5}
}~,
\label{c41SI}
\\
c_{4,2}^{\myrm SI} =&~  { -\frac{481}{1152}
+\frac{971}{1152}  \, \zeta_{3}
-\frac{295}{576}  \, \zeta_{5}
+\frac{11}{32}  \, \zeta_3^2
}~,
\label{c42SI}
\\
c_{4,3}^{\myrm SI} =&~ {
\frac{119}{1152}
-\frac{67}{288}  \, \zeta_{3}
+\frac{35}{144}  \, \zeta_{5}
-\frac{1}{8}  \, \zeta_3^2
}~.
\label{c43SI}
\end{align}

% \ice{
% The polarized Bjorken sum rule  reads
% \begin{equation}
% Bjp(Q^2)=\int_0^1[g_1^{ep}(x,Q^2)-g_1^{en}(x,Q^2)]dx=\frac{1}{6}
% |\frac{g_A}{g_V}|C^{Bjp}(a_s)
% \end{equation}
% Coefficient function $C^{Bjp}(a_s)$ is  fixed by OPE
% of two non-singlet vector
% currents (up to power suppressed corrections)
% \begin{equation}
% i\int T{V_{\alpha}^{a}(x)V_{\beta}^{b}(0)}e^{iqx}dx|_{q^2\rightarrow{\infty}}
% \approx C_{\alpha\beta\rho}^{Q,abc}A_{\rho}^{c}(0)+\dots
% \label{7}
% \end{equation}
% where
% \begin{equation}
% C_{\alpha\beta\rho}^{Q,abc}\sim
% id^{abc}\epsilon_{\alpha\beta\rho\sigma}
% \frac{q^{\sigma}}{Q^2}C^{Bjp}(a_s)
% \nonumberb
% \end{equation}
% and $Q^2=-q^2$
% }

Using  the input from eqs. (\ref{d41}-\ref{d43}) and (\ref{c31SI}-\ref{c43SI}), 
the validity of the GCR can now be investigated. 
In fact, there exist two of them \cite{Adler:1973kz,Broadhurst:1993ru},
one  involving  the non-singlet parts only
and one involving also a singlet piece:
\begin{eqnarray}
D^{\myrm NS}(a_s)\, C^{Bjp}(a_s) &=& d_R \left[ 1+ \frac{\beta(a_s)}{a_s}\, K^{\myrm NS}(a_S)
\right]~,
\label{gCrewtherNS}
\\
  K^{\myrm NS}(a_s) &=&
a_s\,K^{\myrm NS}_1 
+ a_s^2\,K^{\myrm NS}_2 +a_s^3\,K^{\myrm NS}_3
+ \dots
\nonumber
\end{eqnarray}
and 
\begin{eqnarray}
{ \displaystyle D(a_s)\,  C^{GLS}(a_s)}
\label{gCrewtherFull}
&=& 
{d_R \, n_f} \left[ 1 + 
\frac{\beta(a_s)}{a_s}\, K(a_s)
\right]
{},
\\
K(a_s) &=&
a_s\,K_1 
+ a_s^2\,K_2 +a_s^3\,K_3
+ \dots
\nonumber
\end{eqnarray}
Here $\beta(a_s) = \mu^2\,\frac{\mathrm{d}}{\mathrm{d} \mu^2}
a_s(\mu) =-\sum_{i \ge 0} \beta_i a_s^{i+2}$  is the QCD
$\beta$-function with its first term $ \beta_0 = \frac{11}{12}\, C_A -\frac{T}{3}\,n_f$. 
The term proportional to the $\beta$-function describes the deviation from
the limit of exact conformal invariance, with the deviations starting in order $\alpha_s^2$.

Relation (\ref{gCrewtherNS}) has been studied in detail in \cite{PhysRevLett.104.132004},
where its  validity at order $\alpha_s^4$ was demonstrated  
(a  detailed  discussion at orders $\alpha_s^2 $ and $\alpha_s^3$
can be found in  \cite{Broadhurst:1993ru}).

Let us consider now eq. (\ref{gCrewtherFull}).
Combining  
eqs.~(\ref{D:decomp},\ref{GLSdecomp},\ref{chiral:inv}) and (\ref{gCrewtherNS}) leads  
to the following relations between coefficients $K_i^{\myrm NS}$ and $K_i$:
\begin{eqnarray}
K_1 &=& K_1^{\myrm NS}, \ \ K_2 = K_2^{\myrm NS}~,
\\
 K_3 &=& K^{\myrm NS}_3 + K_3^{\myrm SI}~,
\\
K_3^{\myrm SI} &=& k^{\myrm SI}_{3,1} \,\, n_f\,\frac{d^{abc}\,d^{abc}}{d_R}~,
\end{eqnarray}
with  $k^{\myrm SI}_{3,1}$ being a numerical parameter.
 
Thus, we conclude that eq. (\ref{gCrewtherFull}) puts $3-1 = 2$
constraints between two triplets of (purely numerical) parameters $\{
d^{\myrm SI}_{4,1}, d^{\myrm SI}_{4,2}, d^{\myrm SI}_{4,3}\}$ and $\{ c^{\myrm SI}_{4,1},
c^{\myrm SI}_{4,2}, c^{\myrm SI}_{4,3}\}$ appearing in eqs. (\ref{SI:d3:d4}) and (\ref{c4SI}) and
completely describing the order $\alpha_s^4$ singlet contributions to
the Adler function and the Gross-Llewellyn Smith sum rule
respectively.

The solution of the constraints and eqs.~(\ref{c41SI}-\ref{c43SI}) 
produces  the following relations for  $d_4^{\myrm SI}$:
\begin{eqnarray} d_{4,1}^{\myrm SI} &=& -\frac{3}{2}c_{3,1}^{\myrm SI} - c_{4,1}^{\myrm SI}
 =
 -\frac{13}{64} - \frac{\zeta_3}{4}  + \frac{5\zeta_5}{8} ~,
\\
d_{4,2}^{\myrm SI} &=& - c_{4,2}^{\myrm SI}  -  \frac{11}{12}\, k_{3,1}^{\myrm SI}~,
\\
 d_{4,3}^{\myrm SI} &=& - c_{4,3}^{\myrm SI}  +  \frac{1}{3} \,k_{3,1}^{\myrm SI} ~,
\end{eqnarray}
whose validity is indeed confirmed by the explicit calculations. As a
result the remaining unknown $k^{\myrm SI}_{3,1}$ is fixed as:
% \ice{
% K3SI = -179/384 + (25*z3)/48 - (5*z5)/24
% 
% \begin{equation}
% k^{\myrm SI}_{3,1} = -\frac{179}{1152} + \frac{254}{1443} \, \zeta_3 - \frac{10}{72}\, \zeta_5
% {}????.
% \end{equation}
% }
\begin{equation}
k^{\myrm SI}_{3,1} = -\frac{179}{384} + \frac{25}{48} \, \zeta_3 - \frac{5}{24}\, \zeta_5~.
\end{equation}

% \ice{
% \begin{eqnarray}
% &{}& \left(D^{\myrm NS} + \frac{n_f}{d_R}\,D^{\myrm SI}\right) C^{GLS} 
% =
% \\
% &{}&
%  1+ \frac{\beta(a_s)}{a_s}\, K(a_s),
% \\
% &{}& K^{\myrm NS}(a_s) =
% K^{\myrm NS}_0 +
% a_s\,K^{\myrm NS}_1 
% \nonumber
% \\
% && \hspace{1.7cm} + a_s^2\,K^{\myrm NS}_2 +a_s^3\,K^{\myrm NS}_3
% + \dots
% \label{gCrewtherFull}
% \end{eqnarray}
% }

\section{Conclusion}

We have analytically computed coeffcients of  all three colour 
structures  contributing to the singlet part of the Adler function in massless QCD at ${\cal O}(\alpha_s^4)$. 
We have  checked that all constraints on these coefficients derived previously in \cite{Baikov:2010iw} 
on the base of the GCR are  really fulfilled. This is  an important cross-check of 
our calculations of $D^{\myrm SI}$, $C^{\myrm SI}$ and  the very GCR.

The calculations  has been performed
on a SGI ALTIX 24-node IB-interconnected cluster of 8-cores Xeon
computers 
using  parallel  MPI-based \cite{Tentyukov:2004hz} as well as thread-based 
\cite{Tentyukov:2007mu} versions  of FORM
\cite{Vermaseren:2000nd}.  For the evaluation of color factors we have used the FORM program {\em COLOR}
\cite{vanRitbergen:1998pn}. The diagrams have been generated with QGRAF \cite{Nogueira:1991ex}.
The figures have been drawn with the  the help of
Axodraw \cite{Vermaseren:1994je} and JaxoDraw  \cite{Binosi:2003yf}.

This work was supported by the Deutsche Forschungsgemeinschaft in the
Sonderforschungsbereich/Transregio SFB/TR-9 ``Computational Particle
Physics'' and  by RFBR grants  11-02-01196, 10-02-00525.

\providecommand{\href}[2]{#2}\begingroup\raggedright\endgroup

\end{document}